\documentclass[conference]{IEEEtran}
\IEEEoverridecommandlockouts

\usepackage{amssymb}
\usepackage[cmex10]{amsmath}
\usepackage{stfloats}
\usepackage{graphicx}
\usepackage{subfigure}
\usepackage{tabularx}
\usepackage{epsfig,epsf,color,balance,cite}
\usepackage{verbatim}
\usepackage{url}
\usepackage{bm}
\usepackage{booktabs}
\usepackage{diagbox}

\usepackage{algorithm}
\usepackage{algorithmic}

\usepackage{color}
\definecolor{myc1}{rgb}{0,0,0}

\begin{document}
\title{Fair Resource Allocation in UAV-based Semantic Communication System with Fluid Antenna}
\author{Liang Siyun,  
            Chen Zhu,
            Zhaohui Yang,
            Changsheng You,
            Dusit Niyato, \IEEEmembership{Fellow, IEEE,}\\
            Kai-Kit Wong, \IEEEmembership{Fellow, IEEE,}
           and Zhaoyang Zhang, \IEEEmembership{Senior Member, IEEE}

\thanks{S. Liang, Z. Yang, and Z. Zhang are with the College of Information Science and Electronic Engineering, Zhejiang University, and also with Zhejiang Provincial Key Laboratory of Info. Proc., Commun. \& Netw. (IPCAN), Hangzhou, 310027, China (e-mails: \{siyun\_liang, yang\_zhaohui, zhzy\}@zju.edu.cn).}
\thanks{C. Zhu is with Polytechnic Institute, Zhejiang University, Hangzhou, Zhejiang,310015, China (e-mail: zhuc@zju.edu.cn).}
\thanks{C. You is with Department of Electronic and Electrical Engineering Southern University of Science and Technology (SUSTech) Shenzhen, 518055, China (e-mail: youcs@sustech.edu.cn).}
\thanks{Dusit Niyato is with School of Computer Science and Engineering, Nanyang Technological University, 639798, Singapore (e-mail: dniyato@ntu.edu.sg).}
\thanks{Kai-Kit Wong is with the Department of Electronic and Electrical Engineering, University College London, WC1E 7JE London, U.K., and also with the Yonsei Frontier Laboratory, Yonsei University, Seoul 03722, South Korea (e-mail: kai-kit.wong@ucl.ac.uk).}
}
\maketitle
\begin{abstract}
    In this paper, the problem of maximization of the minimum equivalent rate in a unmanned-aerial-vehicle (UAV)-based multi-user semantic communication system is investigated. In the considered model, a multi-antenna UAV employs semantic extraction techniques to compress the data ready to be sent to the users, which are equipped with fluid antennas. Our aim is to jointly optimize the trajectory of the UAV, the transmit beamforming and the semantic compression rate at the UAV, as well as the selection of activated ports in fluid antenna system (FAS), to maximize the minimum  equivalent transmission rate among all user. An alternating algorithm is designed to solve the problem. Simulation results validate the effectiveness of the proposed algorithm.
\end{abstract}
\section{Introduction}

Applications of unmanned aerial vehicle (UAV) in communication networks have attracted a lot of attention from both academia and industries. Benefited from its high motility, and broad range of activity, UAV is a tremendous solution to complex communication scenarios. A significant advantage of UAV application in communication systems is that UAVs' mobility allows them to reach for line-of-sight (LoS) channels with better quality in rich-obstruction environment \cite{8337920}, which enables UAV-aided networks to overcome challenges resulted from blockage in conventional communication systems. Besides, the deployment of UAV is simple and low-costed, making it possible to employ in various scenarios. From data collection in Internet of things (IoT) \cite{9779853} to energy harvesting\cite{8941314} and computing systems \cite{9738797}, the applications of UAV are diversified and flexible. 

Traditional antennas are gradually unable to meet the rigorous requirements of novel communication systems. Consequently, there is a clear necessity for the development of efficient antennas. Fluid antenna is one of the most high-profile technique in this aspect. By reconfiguring the antennas' shape, position and beamforming, fluid antenna has the potential to flexibly adjust the antennas' features according to a given channel state under the control of specific software. Wong et.al analyzed the potential of surpassing other antenna systems in \cite{9131873}, and continued to study the achievable performance of a single fluid antenna system (FAS) in \cite{9264694}, which also presents the first clear definition of FAS. This topic has received wide focus and multiple studies have been conducted to further investigate its application in different wireless communication systems\cite{9539785,10318061,Chen2024JointBA}. Combining FAS, the adaptability of UAV-enabled communication systems is further enhanced. In \cite{10750660}, the authors proposed a UAV relay-aided fluid antenna system with non-orthogonal multiple access (NOMA), where UAV serves as a relay between the users equipped with FAS and the base station (BS).

Physical layer improvements lays the foundation of novel transmission system, while new paradigms are aroused as well. As a new paradigm, the concept of semantic communication is first brought up by Shannon and Weaver.  By extracting the context-related information in the transmitting data, semantic communication guarantees the topic of the information being conveyed correctly.  As for the semantic expression, one valid scheme is knowledge graph. In a knowledge graph, the entities and the relation between two entities are represented as semantic triples. Usage of knowledge graphs is common in semantic communication systems \cite{9416312}. To further exploit the potential of knowledge graph, probability graph is put forward, attaching the corresponding probability of a certain relation occurring between given entities. As a valid form of semantic expression, it can be obtained by performing semantic extraction. But the incidental power consumption address the importance of reasonable resource allocation in this scheme\cite{10915662}.

However, there is still a lack of introducing both fluid antenna and UAV to semantic communication systems. In light of furthering enhance the adaptability of semantic communication systems, we have made the following contributions in this paper:
\begin{itemize}
    \item We consider a downlink system where a multi-antenna UAV transmits semantic information to multiple users with FAS. There are multiple ports in the considered FAS and they can be activated at any given time. Furthermore, we formulate an optimization problem with the objective of maximizing the minimum equivalent rate between the UAV and all users.
    \item To address the formulated max-min problem, we leverage alternative optimization (AO) algorithm and Dinkelbach’s transform to maximize the achievable rate for each user at a given time. Then, we use a modified ant colony algorithm to solve the trajectory planning problem of the UAV.
    \item Numerical results are provided to prove the effectiveness of the proposed algorithm.
\end{itemize}
\section{System Model}
Consider a downlink UAV-enabled communication system, in which a UAV and the $K$ users have access to shared probability graphs. The UAV is equipped with $N$ fixed-position antennas and all users are equipped with an FAS. They can be activated at any given moment. The position of users are denoted by $\boldsymbol{w}=[\boldsymbol{w}_1,\ldots,\boldsymbol{w}_k,\ldots,\boldsymbol{w}_K]$, where $\boldsymbol{w}_k=[x_k,y_k],\forall k\in \mathcal{K}$. For simplicity, we assume that all users located on the ground plane and the UAV flies at a certain height $H$.
\subsection{Semantic Compression Model}
In the proposed system, the UAV performs semantic extraction on the transmitted data $\mathcal{X}_k$ for the $k$-th user via its local probability graph and send the data after compression $\mathcal{Y}_k$ to the $k$-th user. Semantic compression ratio $\rho_k=\frac{\mathrm{size}(\mathcal{Y})}{\mathrm{size}(\mathcal{X}_k)}$, is introduced as a metric of the effect of semantic compression, where $\mathrm{size}(\cdot)$ is a function measuring the size of the data.

Semantic compression requires extra computation power. The computational load for the considered system is formulated as:
\begin{equation}\label{eq:gn}
    c\left(\rho\right)=\left\{\begin{array}{l}
        A_1\rho +B_1, D_1\leq \rho \leq 1, \\
        A_2\rho +B_2, D_2\leq \rho < D_1, \\
        \vdots \\
        A_S\rho +B_S, D_S\leq \rho < D_{S-1},
    \end{array}\right.
\end{equation}
where $0>A_1>A_2>\cdots>A_S$ are the slopes of segments in $c\left(\rho\right)$. $B_s$ is the intercept and $D_s$ is the partition between segments. \eqref{eq:gn} incorporates that a reduction in the compression rate $\rho$ inevitably consumes more power. 

With \eqref{eq:gn}, we can obtain the compression power for the semantic extraction as 
\begin{equation}\label{eq:pc_}
    P_{\rm com}(\rho_k[c]) = c(\rho_k[c])p_0,
\end{equation}
where $p_0$ stands for the positive computation power coefficient, and $c$ is the current time period ($(\rho_k[c]$ denotes the semantic compression ratio for user $k$ at the $c$-th time period). Details of time period will be presented in subsection \ref{sec:t}.

\subsection{Antenna Location Model} 
\subsubsection{BS Antenna Array}
 The transmit antennas are distributed uniformly perpendicular to the ground, while the distance between adjacent antennas is $d_{BS}$. Number the antennas from 1 to $N$ from bottom to top, then the $n$-th antenna's three dimensional (3D) coordinate can be expressed as $[x_{U},y_{U},H+y_{{\rm BS},n}]$, where $x_{U},y_{U}$ is the position of UAV, and $y_{{\rm BS},n}$ is obtained as:
\begin{equation}
    y_{{\rm BS},n}=\frac{2(n-1)-N+1}{2} d_{BS}.
\end{equation}
\subsubsection{Ports in FAS at User Side}
In this paper, $M$ ports of the FAS are aligned evenly perpendicular to the ground, with a distance of $d_{U}$ between two neighbouring ports. The number of activated ports is denoted by $m_0$, which satisfies $1\leq m_0\leq M$. The indices of the selected activated ports for the $c$-period are represented as $\boldsymbol{r}_k[c]=[r_{k,1}, \cdots, r_{k,m}, \cdots, r_{k,m_0}]^T \in \mathbb{Z}^{m_0\times 1}$, where $r_{k,m} \in \{1,2,\ldots,M\},\;r_{k,1}<r_{k,2}<\cdots<r_{k,m_0}$. We number the ports in a manner that ports with larger indices are more distant from the ground. So the 3D coordinate of the $r_{k,m}$-th port is $\boldsymbol{p}_r(k,m)=[x_k,y_k,H+y_{k,m}]$, in which 
$[x_k,y_k]$ is the horizontal position of the $k$-th user, and $y_{k,m}$ is given by:
\begin{equation}
    y_{k,m}=\frac{2(r_{k,m}-1)-M_k+1}{2} d_{U}.
\end{equation}
\subsection{Signal Transmission Model}
 The transmit signal is denoted by $\boldsymbol{x} \in \mathbb{C}^{N\times1}\sim\mathcal{C}\mathcal{N}(\boldsymbol{0},\boldsymbol{Q})$ where $\boldsymbol{Q}\in\mathbb{C}^{N\times N}$ represents the transmit covariance matrix.  The power consumption of the UAV is $\mathrm{tr}(\boldsymbol{Q})$. Taking computation power \eqref{eq:pc_} into consideration, the total power constraint is formulated as:
\begin{equation}
    P_{\rm com}(\rho_k[c])+\mathrm{tr}(\boldsymbol{Q})\leq P_{\max},
\end{equation}
where $P_{\max}$ stands for the total power that can be allocated to transmission and computation. 

The received signal is expressed as:
\begin{equation}
\boldsymbol{y}_k(\boldsymbol{r}_k)=\boldsymbol{G}_k(\boldsymbol{r}_k)\boldsymbol{x}+\boldsymbol{\omega}_k,
\end{equation}
where $\boldsymbol{G}_k(\boldsymbol{r}_k)\in \mathbb{C}^{m_0\times N}$ is the channel coefficient matrix from the transmit antennas to the activated ports in the FAS, and $\boldsymbol{\omega}_k\in\mathbb{C}^{m_0\times 1}\sim \mathcal{C}\mathcal{N}(\boldsymbol{0},\sigma^2\boldsymbol{\mathrm{I}}_{m_0})$ is the complex additive white Gaussian noise with variance $\sigma^2$ for each element in $\boldsymbol{w}_k$.

We assume the channels between the UAV and users are LoS channels. From the established shared coordinate, the distance between the origin of the transmitting antennas and the $m$-th port is written as
\begin{equation}
L_k(m)=\sqrt{(H-y_{k,m})^2+\Vert \boldsymbol{q}[c]-\boldsymbol{w}_k\Vert^2},
\end{equation}
where $\boldsymbol{q}[c]$ is the horizontal position of the UAV in the $c$-th time period. Details of $\boldsymbol{q}$ will be introduced in the next subsection.

Besides, we denote the the angle of departure (AoD) of the $m$-th port of the $k$-th user as $\theta_k(m)$. The sine value of $\theta_k(m)$ is formulated as 
\begin{equation}
\sin\theta_k(m)=\frac{H-y_{k,m}}{L_k(m)}.
\end{equation}

The response coefficient of a specific antenna can be obtained using its propagation path distance difference to a port between it and the origin, which is given as
\begin{align}
    &d_k(n,m)= \nonumber\\
    &\sqrt{{L_k^2(m)}+(y_{{\rm BS},n})^2+2L_k(m) y_{{\rm BS},n}\sin\left({\theta_k(m)}\right)}-L_k(m).
\end{align}
where $n$ denotes the index of the transmit antenna, and $m$ is the index of the receive activated port.
Since we know that the carrier frequency $\lambda$, the signal phase difference between the $n$-th transmit antenna and the origin to the $m$-th port is $2\pi d_k(n,m)/\lambda$. Hence, the transmit field response vector can be written as
\begin{equation}
\boldsymbol{g}_k(n)\triangleq\left[h_{n,1}e^{j\frac{2\pi}{\lambda}d_k(n,1)}, \ldots, h_{n,m_0}e^{j\frac{2\pi}{\lambda}d_k(n,m_0)}\right]^T\in \mathbb{C}^{m_0\times1}.
\end{equation}
where $h_{n,m}$ denotes the channel coefficient for the channel from the $n$-th static antenna to the $m$-th port. It is calculated by $h_{n,m}=\frac{h_0}{\sqrt{(H+y_{BS,n}-y_{k,m})^2+\Vert \boldsymbol{q}[c]-\boldsymbol{w}_k\Vert^2}}$ and $h_0$ is a reference channel coefficient at distance of 1m.

Furthermore, the field response of all the $N$ transmit antennas can be obtained in matrix form as follows:
\begin{equation}
\boldsymbol{G}_k\triangleq[\boldsymbol{g}_k(1), \boldsymbol{g}_k(2), \ldots, \boldsymbol{g}_k(N)]\in\mathbb{C}^{m_0\times N}.
\end{equation}

Knowing the field response and the semantic compression ratio, we could obtain the achievable rate for the $k$-th user:
\begin{equation}
R_k=\frac{1}{\rho_k[c]}\log \det\left(\boldsymbol{\mathrm{I}}_{m_0}+\frac{1}{\sigma^2}\boldsymbol{G}_k\boldsymbol{Q}\boldsymbol{G}_k^H\right), \label{eq:rkc}
\end{equation}
where the transmission rate is divided by a semantic compression ratio $\rho_k[c]$ of user $k$ at time period $c$. Note that \eqref{eq:rkc} only involves the potential transmission rate between the UAV and the user instead of the real one, since the UAV communicates with merely one user in a single time slot. The user-UAV association model will be discussed in the next subsection.
\subsection{User Location Uncertainty}
Users adopt GPS (Global Positioning System) to locate themselves and the UAV sends data according to the users' positions acquired. However, the locating accuracy of GPS is limited, or affected by natural circumstances. Considering the bias of positioning $\Delta\boldsymbol{w}_k=[\Delta x_k, \Delta y_k]$, the $k$-th user's location can be written as
\begin{equation}
    \boldsymbol{w}_k^\prime=\boldsymbol{w}_k+\Delta\boldsymbol{w}_k.
\end{equation}

We define set $\Phi_k$ to collect the uncertainty of the location of the $k$-th user. $\Phi_k$ is given as
\begin{equation}
    \Phi_k=
    \{\boldsymbol{w}_k^\prime\in\mathbb{R}|\Delta\boldsymbol{w}_k\Delta\boldsymbol{w}_k^{\rm T}\leq V\},
\end{equation}
where $V$ denotes the bounded magnitude radius of the region the $k$-th user may locate, and is dependent on the location accuracy.
\subsection{UAV Communication System}\label{sec:t}
The total flight time for UAV is given as $T$ and the UAV serves users via periodic time-division multiple access (TDMA) mode. Divide the flight period into $C$ time slots $\mathcal{C}=\{1,2,\ldots, C\}$. The horizontal coordinate of the UAV in the $c$-th flight period is $\boldsymbol{q}[c]=(x_U(c),y_U(c))$, subjecting to $\boldsymbol{q}[1] = \boldsymbol{q}[C]$, which represents that the UAV flies back to the initial position as the end of its trajectory.

We consider that UAV only serves one user at any given time slot, and introduce $a_k[c]\in \{0,1\}$ to represent the existence of communication between the $k$-th user and UAV. If $a_k[c]=1$, the UAV is serving the $k$-th user at the $c$-th time slot. Otherwise $a_k[c]=0$. For any given time slot $c$, the UAV only keeps in link with one user, so $\sum_{k=1}^{K}a_{k}[c]\leq 1$. By multiplying $\boldsymbol{a}$, we have the equivalent rate for the $k$-th user at the $c$-th time slot as
\begin{align}
R^\prime_k[c]&=a_k[c]R_k[c],\\
a_k[c] &\in \{0,1\},\forall c \in  \mathcal{C},\\
\sum_{k=1}^{K}a_{k}[c] &\leq 1, \forall c \in \mathcal{C}.
\end{align}

 Finally, we define the lower bound of all users' equivalent rate during the whole transmission as
\begin{equation}
     \gamma= \min_{\forall k \in \mathcal{K}} \sum_{c=1}^{C}R^{\prime}_{k}[c].
\end{equation}
\subsection{Problem Formulation}
In this paper, we aim at maximizing the lower bound of all users' equivalent rate with respect to the UAV's trajectory, user association, transmit covariance matrix, semantic compression ratio and the port selection in FAS. The problem can be formulated as:
\begin{subequations}
    \begin{align}
        \max_{\boldsymbol{Q},\boldsymbol{r},\boldsymbol{q},\boldsymbol{a},\boldsymbol{\rho}} \quad& \gamma, \label{eq:gamma}
         \\
    \textrm{s.t.}\quad 
    &\sum_{c=1}^{C}R^{\prime}_{k}[c] \geq \gamma, \forall k, \\
    &\boldsymbol{q}[1]=\boldsymbol{q}[C] \label{eq:return}\\
    &\Vert \boldsymbol{q}[c+1]-\boldsymbol{q}[c] \Vert \leq d_{\max} ,\forall c \in \mathcal{C}, \label{eq:c3}\\
    &r_{k,m}\in \{1,\ldots,M\},\label{eq:c4}\\
    &r_{k,1}< r_{k,2}< \cdots < r_{k,m_0} \label{eq:c5}\\
    &\sum_{k=1}^{K} a_{k}[c] \leq 1, \forall c \in \mathcal{C}, \label{eq:c6}\\
    &a_{k}[c] \in \{0,1\}, \forall k,c, \label{eq:c7}\\
&\mathrm{tr}(\boldsymbol{Q}[c])+P_{\rm com}(\rho_k[c])\leq P_{\rm max},\forall c \in \mathcal{C},\label{eq:proc1}\\
& \rho_k[c] \geq 0,\forall c \in \mathcal{C},\label{eq:proc2}\\
&\boldsymbol{Q}[c]\succeq \boldsymbol{0},\forall c \in \mathcal{C}.\label{eq:proc3}
    \end{align}
\end{subequations}

The constraints \eqref{eq:return} and \eqref{eq:c3} guarantee a legal trajectory of the UAV, while the constraints \eqref{eq:c4} and \eqref{eq:c5} represent the limitation in selecting activated ports. For the constraints \eqref{eq:c6} and \eqref{eq:c7}, they reflect the nature of TDMA. The constraint \eqref{eq:proc1} ensures the power consumption is not exceeded, and the constraint \eqref{eq:proc2} guarantees that the semantic compression ratio is legal. The last constraint \eqref{eq:proc3} let the transmission power be non-negative. Apparently, the problem is non-convex and hard to solve directly.
\section{Algorithm Design}
\subsection{Optimization for Achievable Rate}

It is clear that only when each user's achievable rate is maximized, will the minimum equivalent rate be maximized. In this part of optimization, we aim at maximizing the achievable rate at a given time slot $c$ for a single user $k$.

\subsubsection{Optimization for Transmit Covariance Matrix and Semantic Compression Ratio}
Transfer problem (17) into:
\begin{subequations}
    \begin{align}
        \max_{\boldsymbol{Q}[c],\rho_k[c]} \quad& \frac{1}{\rho_k[c]}\log\det \left(\boldsymbol{\mathrm{I}}_{m_0}+\frac{1}{\sigma^2}\boldsymbol{G}_k\boldsymbol{Q}[c]\boldsymbol{G}_k^H\right), \label{eq:sp1}\\
    \textrm{s.t.}\quad 
&(\ref{eq:proc1})-(\ref{eq:proc3}).\nonumber
    \end{align}
\end{subequations}

It is evident that when the inequality constraint \eqref{eq:proc1} holds with equality, the achievable rate is maximized with a certain $\boldsymbol{Q}[c]$. In this case, \eqref{eq:proc1} is transformed into
\begin{equation}
    P_{\rm com}(\rho_k[c])=P_{\rm max}-\mathrm{tr}(\boldsymbol{Q}[c]) \geq 0.\label{eq:pc}
\end{equation}

From \eqref{eq:gn}, we rewrite the compression ratio $\rho$ with $P_{\rm com}(\rho_k[c])$:

\begin{equation}
    \rho_k[c]=\sum_{s=1}^{S}\frac{P_{\rm com}(\rho_k[c])/p_0-B_s}{A_s} \theta_{s},\label{eq:f}
\end{equation}
where $\sum_{s=1}^{S} \theta_s=1, \theta_s \in \{0,1\}, \forall s \in \{1,2,\ldots,S\}$. $\theta_s$ indicates the segment of compression load function where the current compression ratio lies.

Utilizing \eqref{eq:pc} and \eqref{eq:f}, sub-problem (18) is re-written as:

\begin{subequations}
    \begin{align}
        &\max_{\boldsymbol{Q}[c]} \quad \frac{\log\det \left(\boldsymbol{\mathrm{I}}_{m_k}+\frac{1}{\sigma^2}\boldsymbol{G}_k\boldsymbol{Q}[c]\boldsymbol{G}_k^H\right)}{\sum_{s=1}^{S}\frac{(P_{\rm max}-\mathrm{tr}(\boldsymbol{Q}))/p_0-B_s}{A_s} \theta_{s}} ,\label{eq:sp1_}\\
        \textrm{s.t.}\quad 
&\mathrm{tr}(\boldsymbol{Q}[c])\leq P_{\rm max},\label{eq:sp1fc1}\\
&  P_{\rm max}-\sum_{s=1}^{S} \theta_s (A_s D_{s}+B_s) \nonumber \\
& \leq \mathrm{tr}(\boldsymbol{Q}[c])\leq  P_{\rm max}-\sum_{s=1}^{S} \theta_s(A_s D_{s-1}+B_s) ,\label{eq:segc}\\
&\boldsymbol{Q}[c]\succeq \boldsymbol{0} ,\\
&\sum_{s=1}^{S} \theta_s=1 ,\\
&\theta_s \in \{0,1\}, \forall s \in [1,S],\label{eq:sp1fc6}
    \end{align}
\end{subequations}
where constraint \eqref{eq:segc} is a variant of \eqref{eq:proc1}, and $D_0=1$ denotes compression rate $\rho_k[c]=1$.

We adopt Dinkelbach's transform to solve (21) in an iterative manner. First, we let $\theta_1=1$. A balanced factor $\delta$ is introduced. Denote the numerator of (21) as $p(\boldsymbol{Q}[c])$ and the denominator as $q(\boldsymbol{Q}[c])$, then rewrite the optimization problem (21) as follows:

\begin{subequations}
    \begin{align}
        \max_{\boldsymbol{Q}[c]} \quad& p(\boldsymbol{Q}[c])-\delta q(\boldsymbol{Q}[c]),\label{eq:sp1f}\\
        \textrm{s.t.}\quad 
        &(\ref{eq:sp1fc1})-(\ref{eq:sp1fc6}),\nonumber
    \end{align}
\end{subequations}
where  $\delta=\frac{p(\boldsymbol{Q}^\prime[c])}{q(\boldsymbol{Q}^\prime[c])}$, and $\boldsymbol{Q}^\prime[c]$ is the value of the last transmit matrix obtained. We first find an appropriate value of $\boldsymbol{Q}[c]$ as $\boldsymbol{Q}^{(0)}[c]$, the initial value for updating. Consequently, $\delta^{(0)}$ is determined as $\delta^{(0)}=\frac{p(\boldsymbol{Q}^{(0)}[c])}{q(\boldsymbol{Q}^{(0)}[c])}$. Utilizing $\delta^{(0)}$ in problem (22), problem (22) is transformed into a convex problem, and can be solved by existing convex problem solvers such as CVX. The solution is denoted by $\boldsymbol{Q^{(1)}[c]}$, which will be applied in calculating $\delta^{(1)}$ in the next iteration. By repeating the process until the value of problem (22) converges with a given accuracy $\epsilon_1$, we will obtain a solution $\boldsymbol{Q}[c]$ and $\rho_k[c]$.  After that, we set $\theta_2=1$ and go over the transform and solve (22) with different parameters, until all situations of $\boldsymbol{\theta}$ are considered.
\subsubsection{Optimization for Port Selection}

With transmit covariance matrix $\boldsymbol{Q}[c]$ and compression ratio $\rho_k[c]$ known, the optimization problem can be re-formulated to:
\begin{subequations}
    \begin{align}
        \max_{\boldsymbol{r}_k[c]} \quad &\frac{1}{\rho_k[c]}\log\det \left(\boldsymbol{\mathrm{I}}_{m_k}+\frac{1}{\sigma^2}\boldsymbol{G}_k\boldsymbol{Q}[c]\boldsymbol{G}_k^H\right), \label{eq:r}\\
        \textrm{s.t.}\quad &r_{k,m}\in\{1,\ldots,M\},\\
            &r_{k,1}< r_{k,2}< \cdots < r_{k,m_0}.\label{eq:cr}
    \end{align}
\end{subequations}

It's difficult to enumerate all combinations for $\boldsymbol{r}_k[c]$ in \eqref{eq:r}. We seek to optimize a single term $r_{k,m}$ in $\boldsymbol{r}_k[c]$ instead. With this motivation, we denote the $m$-th row in $\boldsymbol{G}_k$ as
\begin{equation}
    \hat{g}_k(m)\triangleq\left[h_{1,r_{k,m}}e^{j\frac{2\pi}{\lambda}d(1,r_{k,m})}, \ldots, h_{n,r_{k,m}}e^{j\frac{2\pi}{\lambda}d(n,r_{k,m})}\right].
\end{equation}

With $r_{k,m}$ varying, the corresponding $\hat{g}_k(m)$ changes. So, by trying the possible value of $r_m$ according to the constraint \eqref{eq:cr} and computing the value of \eqref{eq:r}, we update $r_{k,m}$ when the new value of \eqref{eq:r} is larger than the current one, which ensures that the value of \eqref{eq:r} is non-strictly increasing within the optimization process. The details for optimizing achievable rate is shown in Algorithm \ref{al:1}.
\begin{algorithm}
    \caption{Optimization for Achievable Rate for User $k$ at time period $c$}
    \begin{algorithmic}\label{al:1}
        \STATE Initialize iteration index $i=0$, tolerance $\epsilon_2$.
        \REPEAT
            \STATE Initialize $\boldsymbol{Q^{(0)}[c]}$
            \STATE Solve problem (18) and obtain $\boldsymbol{Q[c]}$ and $\rho_k[c]$.
            \STATE Solve problem (23) and obtain $\boldsymbol{r}_k[c]$.
            \STATE Calculate the achievable rate $R_k^{(i)}[c]$.
        \UNTIL{$R^{(i)}_k[c]-R^{(i-1)}_k[c]<\epsilon_2$}
    \end{algorithmic}
\end{algorithm}

Algorithm \ref{al:1} includes an alternative optimization which solves semi-defined convex problem repeatedly, and a numerate-like port selection. The time complexity for the alternative optimization using Dinkelbach's transform is $O(\frac{N^{4.5}\log(1/\epsilon_1)}{\epsilon_2})$. In the worst case, we need to go over all possible combinations of activated ports, so the worst-case time complexity for algorithm 1 is $O(\frac{N^{4.5}\log(1/\epsilon_1)}{\epsilon_2}M^{m_0})$. However, in practice, we can initialize $\boldsymbol{r}_k[c]$ in an equational manner (let the activated ports distribute uniformly), which accelerates the convergence, and the convergence speed of Algorithm \ref{al:1} mainly depends on solving the semi-defined convex problem.

\subsection{User Association Optimization}
In this part of optimization, we focus on the optimization for user association $\boldsymbol{a}$.
\begin{subequations}
    \begin{align}
        \max_{\boldsymbol{a}} \quad& \gamma, 
         \label{eq:oa}\\ 
    \textrm{s.t.}\quad 
    &\sum_{k=1}^{K}a_{k}[c] \leq 1, \forall c \in \mathcal{C},\\
    &a_{k}[c] \in \{0,1\}, \forall k,c. \label{eq:ac}
    \end{align}
\end{subequations}

With UAV's trajectory, transmit covariance matrix, semantic compression ratio and port selection determined,  problem (25) is reduced to a binary programming problem. If $C$ and $K$ are small, it is possible to solve problem (25) with existing solvers as Gurobi. However, as the scale of the problem grows, the complexity of solving problem (25) rises significantly. To solve the problem in an acceptable complexity, we relax the constraint \eqref{eq:ac} into 
\begin{equation}
    0\leq a_{k}[c]\leq 1, \forall k,c.\label{eq:akc}
\end{equation}

And problem (27)is transformed into
\begin{subequations}
    \begin{align}
        \max_{\boldsymbol{a}} \quad& \gamma ,
         \label{eq:oa}\\ 
    \textrm{s.t.}\quad 
    &\sum_{k=1}^{K}a_{k}[c] \leq 1, \forall c \in \mathcal{C},\\
    &0\leq a_{k}[c]\leq 1, \forall k,c, \label{eq:ac}
    \end{align}
\end{subequations}
which is a linear programming problem, and can be easily solved. After obtaining the solution $\boldsymbol{a}$, we adopt the rounding method in \cite{wuJointTrajectoryCommunication2018} for a final solution.
\subsection{Trajectory Optimization}

With varying phases of the transmission signal, the optimization of UAV's trajectory is non-convex, and solving it with mathematical method will be complex. In this paper, we use ant colony algorithm to address the problem.

The basic idea of ant colony algorithm is to employ multiple agents acting as ants. They follow a specific hormone-based path-choosing policy, while leaving a certain amount of pheromones on the path they have went through. 

\subsubsection{Initialization}
We divide the searching area into discrete square grids, and formulate a Cartesian coordinate on it. The length of a cell is equal to the maximum moving distance of the UAV $d_{max}$ in a time slot, as a guarantee of constraint \eqref{eq:c3}. During the search, the UAV moves from the centre of a cell to the centre of an adjacent cell.

All cells are assigned with a coordinate $\boldsymbol{u}=(i,j), i\in\{1,2\ldots,I_x\},j\in \{1,2\ldots,I_y\}$, where $I_x$ and $I_y$ are the maximum coordinate indexes for the $x$ and $y$ axis, respectively. Upon these cells, we initialize the pheromone $\tau_{0}$, as \eqref{eq:it} shows.

\begin{equation}
\tau_{\boldsymbol{u}}=\tau_{0},\boldsymbol{u}\in\mathcal{U},\label{eq:it}
\end{equation}
where $\mathcal{U}$ denotes the set of all grid cells. 

\subsubsection{Ant Exploration}
At the beginning of each iteration, we place $N_A$ ants as agents in the starting cell of UAV. In the exploration phase, each ant will search for a path cycling around the searching area and return to the starting point at the end of the trajectory. 

To ensure that the trajectory of each ant satisfies
\eqref{eq:return}, we divide the exploration period into two stages. In the first stage, the current time slot $c$ and the ant's trajectory $q_a$ should be subject to
\begin{equation}
    C-c> \Vert \boldsymbol{q}_a[c]-\boldsymbol{q}_a[1]\Vert_2 . \label{eq:returnc}
\end{equation}

Inequality \eqref{eq:returnc} guarantees that the ants always have enough time to go straight back to the starting cell. In this stage, the probability for an ant to visit an adjacent cell $\boldsymbol{u}$ is given as
\begin{equation}
    p_{\boldsymbol{u}}=\left\{\begin{aligned}
    &\frac{\tau_{\boldsymbol{u}}^\alpha h^\beta(\boldsymbol{u})}{\sum_{ \boldsymbol{i} \in \mathcal{U}_a }\tau_{\boldsymbol{i}}^\alpha h^\beta(\boldsymbol{i})},\boldsymbol{u} \notin \boldsymbol{q}_{ac} \;\rm and\; \mathcal{U}_a \in \boldsymbol{q}_{ac}, \\
    &0,{\rm else}
    \end{aligned}\right.\label{eq:antp}
\end{equation}
where $\mathcal{U}_a$ is the set of all adjacent cells, $\boldsymbol{q}_{ac}$ denotes $\{\boldsymbol{q}_a[1],\ldots,\boldsymbol{q}_a[c]\}$, which is the route the current ant has gone through. By forbidding the ant from moving to an explored cell, we hope the ants explore new cells instead of keeping advancing and returning. However, some ants, especially at the beginning of the algorithm process where there is little pheromone hint, may get into a cell whose  adjacent cells are all explored. Only in this case, we allow the ant to move to an explored cell. $\alpha$ and $\beta$ are preset parameters of the algorithm. They decide the importance of existing pheromone and heuristic information in the path-chosen progress, respectively.  $h(\cdot)$ denotes a heuristic function that offers additional information for the input cell. In this paper, $h(\cdot)$ is given as follow
\begin{equation}
    h(\boldsymbol{u})=\sum_{k=1}^{K} \frac{1}{\Vert {\rm position}(\boldsymbol{u})- \boldsymbol{w}[k] \Vert}, \label{eq:hu}
\end{equation}
where ${\rm position}(\boldsymbol{u})$ is the coordinate of cell $\boldsymbol{u}$ in the universal coordinate system. \eqref{eq:hu} encourages our ants to move towards users.

As $c$ increases, \eqref{eq:returnc} holds with equality, which means the current ant must be on its way to the starting point. Then, the exploration enters the second stage. In this stage, the ant decides its next cell to visit based on
\begin{equation}
    p_{\boldsymbol{u}}=\left\{\begin{aligned}
    &\frac{\tau_{\boldsymbol{u}}^\alpha h^\beta(\boldsymbol{u})}{\sum_{ \boldsymbol{i} \in \mathcal{U}_a }\tau_{\boldsymbol{i}}^\alpha h^\beta(\boldsymbol{i})},\begin{aligned}
            &\Vert {\rm position}(\boldsymbol{u})-\boldsymbol{q}_a[1])\Vert_2<\\
            &\Vert \boldsymbol{q}_a[c]-\boldsymbol{q}_a[1])\Vert_2,\\
        &(\boldsymbol{u} \notin \boldsymbol{q}_{ac} \;\rm and\; \mathcal{U}_a \in \boldsymbol{q}_{ac}),
    \end{aligned} \\
    &0,{\rm else}.
    \end{aligned}\right.\label{eq:antp2}
\end{equation}

The main idea of \eqref{eq:antp2} is similar to \eqref{eq:antp}. The biggest difference is that \eqref{eq:antp2} only chooses cells that are closer to the starting point. If $C-c<2$ and $\boldsymbol{q}_a[c]=\boldsymbol{q}_a[1]$, we simply let ant stay in the starting cell.

\subsubsection{Updating Pheromone Matrix}
After all ants' exploration end, we need to update the current pheromone matrix. First, the existed pheromones is reduced by a volatile rate $v$, shown as follows:
\begin{equation}
    \tau_{\boldsymbol{u}}=(1-v)\tau_{\boldsymbol{u}}. \label{eq:re1}
\end{equation}

Then, for every ant, the trajectory is given. We maximize the achievable rate for one user leveraging Algorithm \ref{al:1}. Then, by addressing problem \eqref{eq:oa}, we can have the optimal minimum  equivalent rate for a specific trajectory, and the reward for a certain trajectory is counted as 
\begin{equation}
    {\rm reward}_a= \min_{\forall k \in \mathcal{K}} \sum_{c=1}^{C}R^{\prime}_{k}[c]\label{eq:reward}.
\end{equation}

Conventional pheromone updating policy in ant colony algorithm will take the reward from every ant into consideration. In this paper, we adopts a modified version of pheromone updating policy, in which only the pheromone of the trajectory with largest reward (calculated from \eqref{eq:reward}) in one round will be selected.
\begin{equation}
    \tau_{\boldsymbol{u}}=\tau_{\boldsymbol{u}}+{\rm reward}_{\rm best}, \boldsymbol{u} \in \{\boldsymbol{q}[1],\ldots,\boldsymbol{q}[C]\}.\label{eq:re2}
\end{equation}

With the new pheromone matrix generated, the next round of ant exploration could start. Details of the ant colony algorithm is presented in Algorithm \ref{al:ant}.
\begin{algorithm}
    \caption{Ant Colony Optimization for UAV Trajectory}
    \begin{algorithmic}\label{al:ant}
        \STATE Input: the number of ants $N_A$, the iteration round $I_r$, initial pheromone concentration $\tau_0$, parameters $\alpha$ and $\beta$, iteration index i=0.
        \STATE Initialize the grids in searching area,  and pheromone $\boldsymbol{\tau}$ by \eqref{eq:it}.
        \WHILE{$i\leq I_r$}
            \FOR{Every ant}
                \STATE Explore and return according to \eqref{eq:antp} and \eqref{eq:antp2} respectively.
            \ENDFOR
            \FOR{Every ant}
                \STATE  Iteratively solving problem (18) and problem (23) and maximizing each user's equivalent rate at a given position.
                \STATE Solve problem (27) and calculate the value of (27) as the reward of the path discovered by the current ant.
            \ENDFOR
            \STATE Find the path with largest reward in all paths explored.
            \STATE Update pheromone according to \eqref{eq:re1} and \eqref{eq:re2}.
        \ENDWHILE
        \STATE Output the best path found during the whole process.
    \end{algorithmic}
\end{algorithm}

With given iteration round $I_r$, number of ants $N_A$ and number of time periods $C$, the time complexity of Algorithm \ref{al:ant} hinges on solving sub problem (27), which is a linear programming problem, and its time complexity varies with the solving method. For simplex method, the worst-case complexity for solving (27) is exponential. But in most cases the algorithm converges with an acceptable speed.
\subsection{Uncertain User Location}
With uncertain user location, the channel gain $h_{n,m}^2$ from the $n$-th antenna of the UAV to the $m$-th port becomes uncertain, and the actual transmission rate for user fluctuates. To ensure the performance of the overall system, we set a lower-bound for transmission rate and finish the optimization with this lower-bound.

Considering the user location uncertainty, the gain of channel between UAV and the $m$-th port of the $k$-th user becomes
\begin{equation}
    h_{n,m}^\prime=\frac{h_0}{\sqrt{(H+y_{BS,n}-y_{k,m})^2+\Vert \boldsymbol{q}[c]-\boldsymbol{w}_k^\prime\Vert^2}}.\label{eq:hhat}
\end{equation}

And the gain in the channel matrix $G$ should be replaced by the $h_{n,m}^\prime$ from \eqref{eq:hhat} in calculation. However, it is impossible for the UAV to obtain the accurate value of $\boldsymbol{w}_k^\prime$. To take the uncertainty into consideration during the optimization, it is necessary to transform \eqref{eq:hhat} into a determined equation.
\section{Simulation Results}
 This section presents the results of simulations conducted on the proposed algorithm. Two additional schemes have been
 conducted as comparison, as described below:
\begin{enumerate}
    \item \textbf{Rectangle:} In this scheme, the UAV's trajectory forms a rectangle lying at the centre of the searching area.
    \item \textbf{Normal-AC:} In this scheme, the pheromone is set on the edges connecting adjacent cells as normal ant colony algorithms applied on route searching or travelling salesman problem (TSP) do. Specifically, $\tau_{i,j}$ represents the pheromone on the path from the $i$-th cell to the $j$-th cell. The iteration rounds and number of ants used each round are the same as scheme `Proposed'. 
\end{enumerate}

In simulation, we set the number of transmit antennas $N=20$, and the total number of ports $M=35$. Number of activated ports $m_0$ is 5. The flight height for UAV 
 $H=30m$, and maximum moving distance in a time slot $d_{max}$ is set to 50m. The carrier frequency $\lambda=4mm$, and distance between adjacent transmit antennas or ports is $\lambda/2$. Maximum power $P_{max}$ is set to $20W$, and the reference channel gain $h_0^2$ is set to -10dBm. The total iteration time for ant colony search is set to 100 and the ants for each round is 50. 
\begin{figure}
    \centering
    \includegraphics[width=0.75\linewidth]{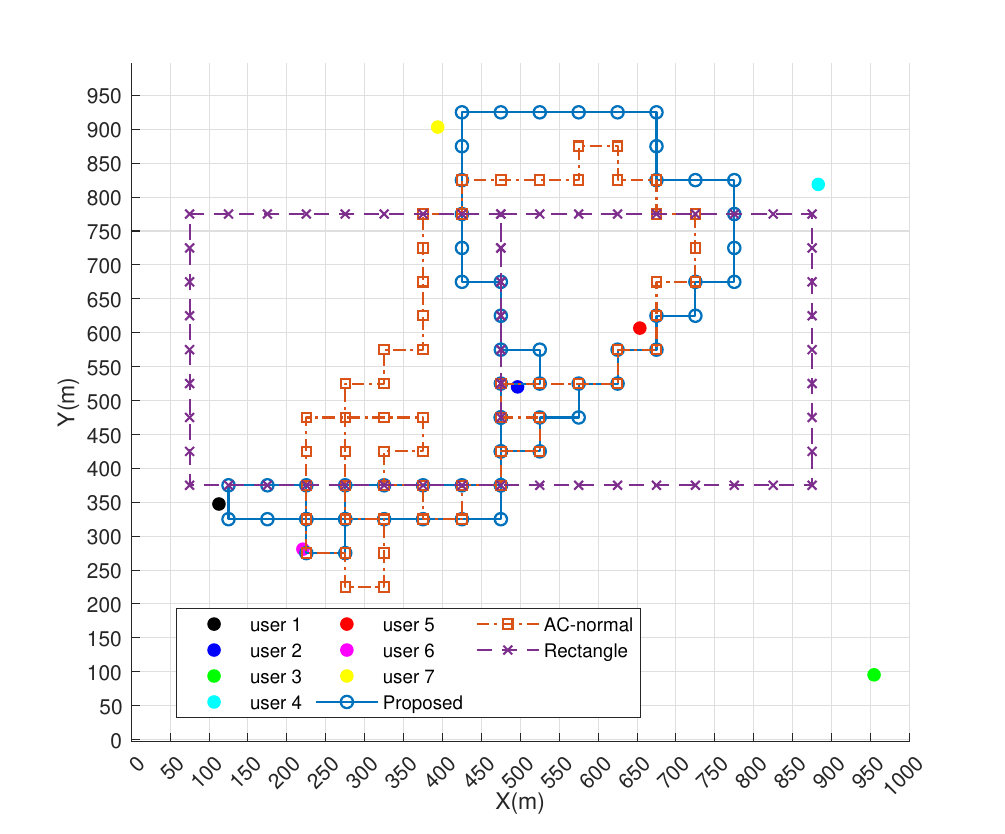}
    \caption{Explored routes of different schemes ($T=60$).}
    \label{fig:1}
\end{figure}
\begin{figure}
    \centering
    \includegraphics[width=0.75\linewidth]{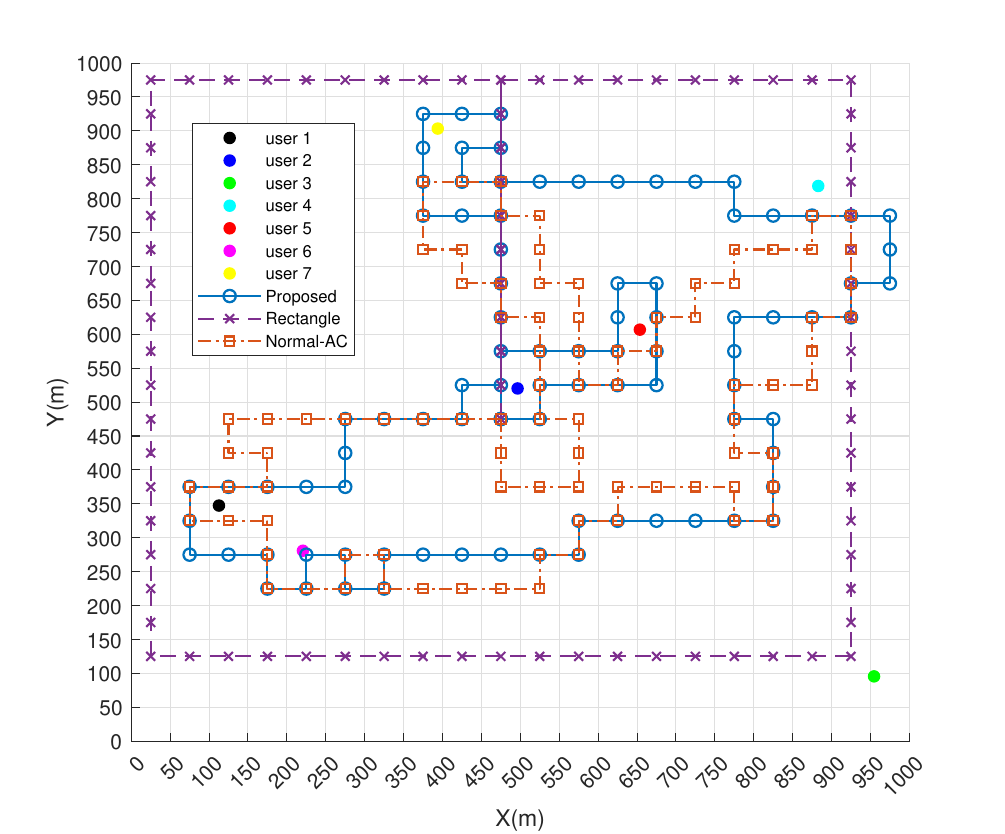}
    \caption{Explored routes of different schemes ($T=90$).}
    \label{fig:2}
\end{figure}
\begin{table}[]
\centering
\caption{Equivalent rate of the 3 schemes conducted with different time periods}
\label{tab:res}
\begin{tabular}{|l|c|c|}
\hline
\diagbox{Scheme}{Equivalent rates}{Periods} & T=60s & T=90s  \\
\hline
Proposed (7 users) & 407.47 & 621.46\\
\hline
Rectangle (7 users) & 385.98 & 572.34  \\   
\hline
Normal-AC (7 users)&384.96& 606.77 \\
\hline
Proposed (10 users)& 289.14 & 441.07\\
\hline
Rectangle (10 users)& 264.28 & 414.35  \\   
\hline
Normal-AC (10 users)& 270.53 & 418.36 \\
\hline
\end{tabular}
\end{table}
As shown in Fig.~\ref{fig:1}, there were 7 users scattering randomly in a square area with a side length of 1 km, and the UAV started at the centre of the area. The coordinate of the starting cell was (475,475), or (10,10) in the grid coordinate. The minimum equivalent rate obtained by 3 different schemes is shown in Table.~\ref{tab:res}. Scheme `Proposed' and scheme `Normal-AC' present significant adaptability to the distribution of users, while scheme `Proposed' displays a lead over the other scheme, showing the effectiveness of the proposed algorithm. 

According to  Table.~\ref{tab:res}, a large number of users will add to the complexity of solving the problem and lead to a strain on communication resources, while scheme `Proposed' shows a greater advantage, indicating its potential to accommodate complex circumstance. 

In Fig.~\ref{fig:2}, the total flight time $T$ is extended to 90 seconds, which leads to a larger decision space comparing to that in Fig.~\ref{fig:1}. It is worth noting that the possible flight distance is not enough to fully cover the 7 users and return. Scheme `Proposed' still presents advantages in accommodating the users' distribution. 
\section{Conclusion}
In this paper, we have formulate a fluid antenna assisted semantic communication system applied on UAV. The BS mounted on the UAV would compress the transmit data leveraging probability graphs before sending them to users equipped with FAS in TDMA mode. In addition, we have formulated a minimum rate maximizing optimization problem, and have adopted an AO algorithm and a modified ant colony algorithm to solve it. Numerical results of simulation have proven the effectiveness of the proposed algorithm.
\bibliographystyle{IEEEtran}
\bibliography{ref}
\end{document}